\documentstyle[prb,aps,twocolumn,graphicx]{revtex}  
\def\CD{ClMePD-DMeDCNQI}        
\def\C{ClMePD}                  
\def\D{DMeDCNQI}                
\def\cm{cm$^{-1}$}               
\begin{document}
\draft

\title{Pressure driven neutral-ionic transition in {\CD}}

\author{Luca Farina and Aldo Brillante}
\address{Dip. di Chimica Fisica ed Inorganica, Universit\`a di
Bologna, Viale Risorgimento 4, 40136-I Bologna, Italy}

\author{Matteo Masino and Alberto Girlando}
\address{Dip. Chimica Generale ed Inorganica, Chimica Analitica e 
Chimica Fisica, Universit\`a di Parma, Parco Area delle Scienze,
43100-I Parma, Italy}

\wideabs{
\maketitle

\begin{abstract}

Application of about 0.8 GPa pressure is sufficient
to induce the neutral-ionic transition in
the mixed stack charge-transfer crystal
2-chloro-5-methyl-$p$-phenylenediamine--2,5-dimethyl-dicyanoquinonediimine
({\CD}). The ionicity increases continuously from $\sim$ 0.35
at ambient conditions to $\sim$ 0.65 when the pressure is
raised up to 2 GPa. Moreover, stack dimerization
begins well before the crossing of the neutral-ionic interface.
The evolution of the transition is similar to what observed
in the temperature induced phase change in the same compound.
A distinguishing feature is represented by the simultaneous
presence of domains of molecules with slightly different ionicities
across the transition pressure. A comparison of the
present example of pressure driven neutral-ionic transition
with the well studied cases of tetrathiafulvalene--chloranil
and of tetrathiafulvalene--2,5-dichloro-$p$-benzoquinone
puts in evidence the remarkably different evolution of the
three transitions.
\end{abstract}
}

\section{Introduction}

Neutral-ionic (NI) phase transitions in mixed stack
charge transfer (CT) crystals\cite{torrance1981} have recently
received renewed interest due to the expectation of unusual physical
properties at the NI borderline. For instance,
exact diagonalization studies have suggested the 
possible existence of a metallic 
state just at the point marking the passage from N to I
ground state of a regular mixed stack chain.\cite{VBopt,VB,asp}
Metallic behavior around this particular point
has been indeed discovered.\cite{saito}
Moreover, a large increase of the dielectric constant,
associated with the collective transfer of charge, 
has been found in proximity of the NI transition.\cite{horiuchi00}
Finally, systems on the I side show
strong tendency towards dimerization, due to the Peierls
instability,\cite{VB} and may therefore display a ferroelectric
ground state.\cite{cailleau1997}
To explore such phenomena, it is useful to investigate CT
crystals that undergo a continuous ionicity change across
the NI borderline. Several years ago our group reported an example
of continuous ionicity change under pressure for
tetrathiafulvalene--2,5-dichloro-$p$-benzoquinone (TTF-DClBQ).\cite{ttf25bq}
More recently, a first example of a continuous NI transition
induced by lowering temperature has been observed for
2-chloro-5-methyl-$p$-phenylenediamine--2,5-dimethyl-dicyanoquinonediimine
({\CD}).\cite{aoki} A detailed characterization of this transition has
shown that there is a sort of feedback mechanism
between the Peierls dimerization transition and the increase of ionicity
at the NI interface.\cite{noi}

In the present paper we investigate the NI phase transition in {\CD}
induced by pressure at room temperature.
We find that the pressure induced phase transition in {\CD} is somewhat
different from the transition induced by temperature.\cite{noi}
Furthermore, in {\CD} the ionicity change across
the NI borderline is practically continuous, as in TTF-DClBQ,\cite{ttf25bq}
but the evolution of the two transitions is quite different.
We also make comparison with the NI phase transition of the 
prototype compound tetrathiafulvalene--chloranil (TTF-CA),
which is known to exhibit a discontinuous NI phase transition
both under pressure\cite{ttfcap,ttfcaj} and under
temperature.\cite{ttfcajcp,ttfcat} 

\section{Experimental}

{\CD} crystals have been prepared by mixing saturated solutions
in dichlorometane followed by slow evaporation of the
solvent.\cite{aoki} 
Infrared (IR) spectra have been recorded with a Nicolet Nexus 470 FTIR on
powdered samples loaded in a diamond-anvil cell (DAC) with nujol or
perfluorocarbons (PFC) as pressure media. High pressure Raman spectra were
obtained with a Renishaw System 1000 Microscope, with excitation from an Ar
ion laser ($\lambda$ = 514.5 nm). A tiny crystal was loaded in the
DAC using nujol as pressure medium. 
The spectral resolution in both IR and Raman spectra is 2 {\cm}.
Pressure calibration was done with the ruby
luminescence technique.\cite{piermarini} 
Accuracy of the pressure reading was about 0.05 GPa in the
pressure range up to 3 GPa. 

\section{Results}

{\CD} crystallizes in the triclinic system, space group P1,
with a=7.463\AA, b=7.504\AA, c=7.191\AA, $\alpha$=91.23$^{\circ}$, 
$\beta$=112.19$^{\circ}$, $\gamma$=96.91$^{\circ}$, and Z=1.~\cite{aoki} 
The electron donor (D) {\C} and the electron acceptor (A)
{\D} form mixed stack columns along the $b$ axis.
The crystal does not have symmetry elements, and the
molecules are located in general positions. Therefore
the stack structure is not constrained by symmetry.

\begin{figure}[ht]
\includegraphics* [scale=0.45]{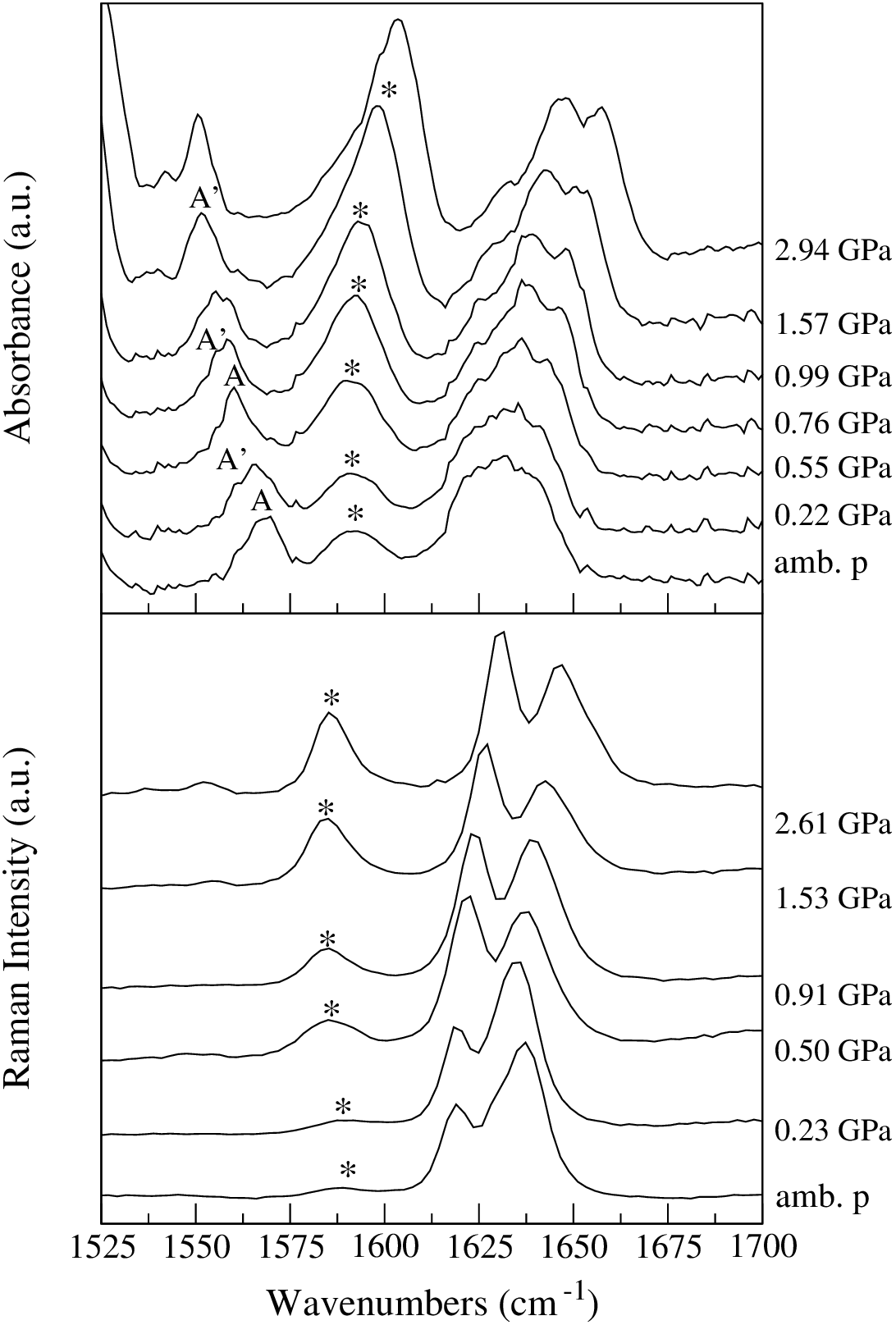}
\protect\vskip0.3truecm
\caption{Pressure dependence of the IR and Raman
spectra of {\CD} in the 1525-1700 {\cm} spectral region. The
pressures are indicated on the right side of the figure.}
\end{figure}

\noindent
Within the accuracy of the X-ray analysis,\cite{aoki}
at room temperature the stack is regular, that is,
each molecule has equal distance (CT integral)
with its two nearest neighbors along the chain. However,
the stack structure presents disorder in the 
relative orientation of the two molecules: the 2- and 5-
substitutional sites of {\C} are considered to be occupied by
chlorine and methyl group with equal probability along the stack.

As it is well known,\cite{ttfcajcp} vibrational
spectroscopy is a very useful tool to study NI
phase transitions of mixed stack CT crystals.
Vibrational frequencies respond to molecular
charge variations, therefore yielding a method
to directly estimate the degree of ionicity $\rho$.
Moreover, the interaction between the CT electrons
and the molecular vibrations ({\it e-mv} coupling)
allows one to discriminate regular from dimerized
stack structure: The stack dimerization is marked
by the appearance in the IR spectra of the (Raman
active) totally symmetric modes, which are
forbidden in the case of regular stack.\cite{40years,bibbia}
The IR intensity of these modes is
related to the extent of dimerization, and can be taken
as the order parameter of the dimerization phase transition.

\begin{figure}[ht]
\includegraphics* [scale=0.45]{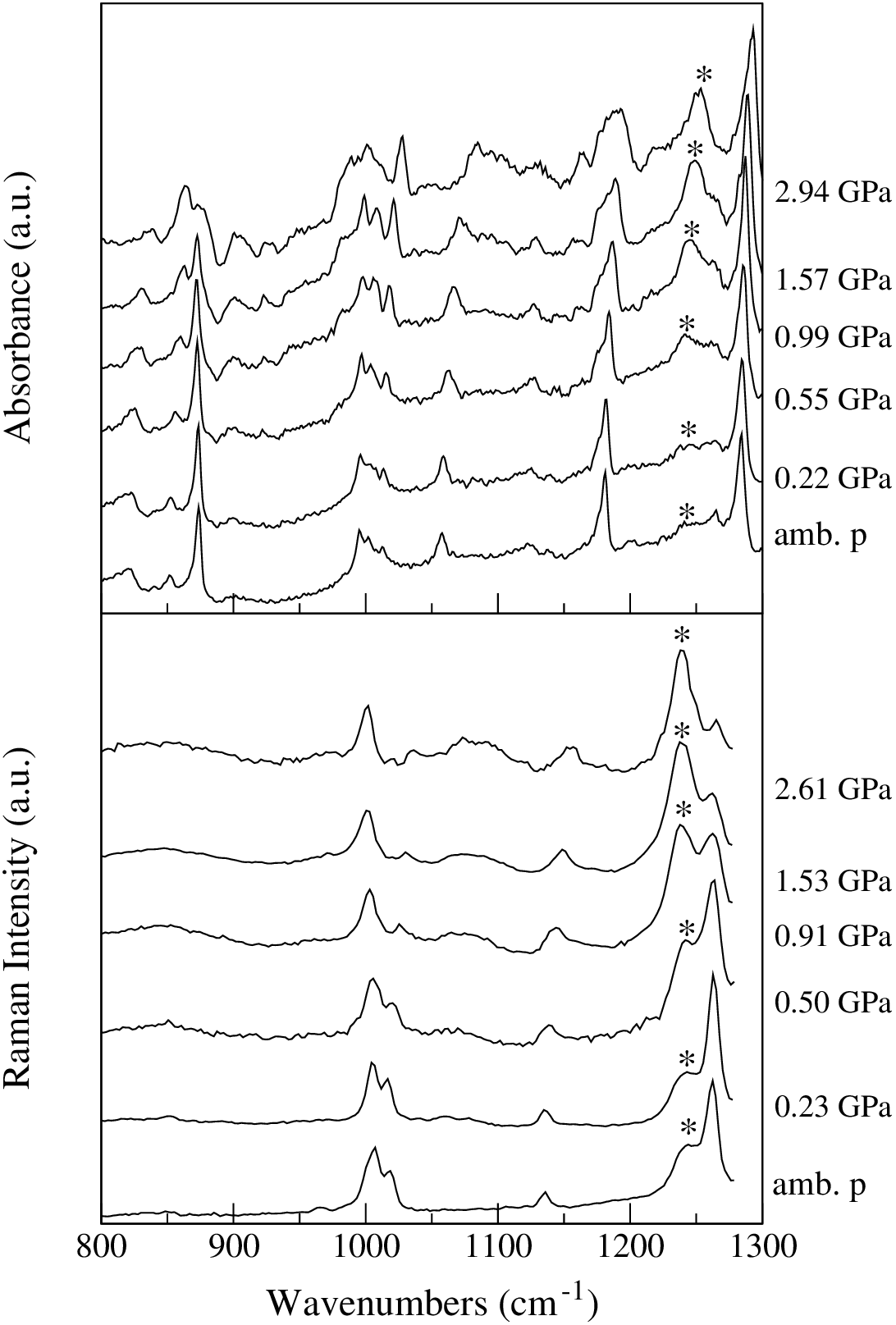}
\protect\vskip0.3truecm
\caption{Pressure dependence of the IR and Raman
spectra of {\CD} in the 800-1300 {\cm} spectral region. The
pressures are indicated on the right side of the figure.}
\end{figure}

In this paper we only deal with the spectral features
directly connected with the phase transition,
as singled out in ref. \onlinecite{noi}, where
a satisfactory and rather complete
interpretation of the IR and Raman spectra has been achieved. 
Figs. 1 and 2 report the IR and Raman spectra
as function of pressure, in the spectral regions 1525-1700
and 800-1300 {\cm}, respectively.
The IR active mode {\D} $b_u \nu_{46}$ mode (1570 {\cm} 
at ambient conditions, marked A in the fig. 1)
has been chosen as the primary probe of ionicity,
since it shows a large ionization frequency
shift (61 {\cm}).\cite{noi}
The other {\D} mode showing a linear dependence on
$\rho$, the $b_u \nu_{47}$, cannot be used 
as a secondary check of the ionicity,\cite{noi} since in 
the lack of polarized IR data we cannot try to disentangle this 
mode in the very complex spectral region 1480-1520 {\cm}.

By increasing the pressure 
the {\D} $b_u \nu_{46}$ (A band) shifts towards lower frequencies
from 1570 to 1550 {\cm}, indicating an increase in the molecular ionicity.
We notice, however, that already at 0.22 GPa a dip 
appears on the low-frequency side of the A band. This
spectral feature (marked A' in fig. 1) gradually 
gains intensity as the pressure increases. 
Between 0.76 and 0.99 GPa its intensity gets comparable to that of 
the A band, and when pressure is raised up to 1.57 GPa the A band
merges into the A' band.                                                  
Since only the {\D} $b_u \nu_{46}$
mode is expected in this frequency region, we interpret
the A and A' bands as due to this mode, in correspondence
with species of different ionicities. Thus the simultaneous 
presence of A and A' bands signals the presence of domains of
slightly different ionicities in the crystal. This kind of behavior,
which would suggest staging in the phase transition, 
is not observed in the temperature induced transition, 
where coexistence of phases at different ionicities can 
safely be excluded.\cite{aoki,noi} However, the behavior
reported in fig. 1 varies on changing sample during different
pressure runs, showing slightly different profiles with respect
to those reported in fig. 1. In any case the structure of
the A band is always rather complex showing in some cases
appreciable intensity of the A' band already at very low pressures.
The problem~ might be~ accounted for the~ different samples
used in IR spectra,
i.e. powdered crystals dispersed in nujol or PFC mulls.
As already observed in the NI transition of TTF-CA,\cite{ttfcaj,ttfcajcp}
defects and imperfections may indeed 
affect the course of the phase transition, and 
measurements on single crystals are needed
to safely ascertain the presence of staging. In fig. 3,
upper panel,  we report the
pressure dependence of the $b_u \nu_{46}$ frequency, taken
as the peak frequency of the most intense among the A, A' bands.

From the pressure evolution of IR spectra in figs. 1 and 2, one
easily recognizes the presence of {\it e-mv} induced
bands, which signal the onset of stack dimerization.
We focus on the two {\it e-mv} induced IR bands at 1587 and 1241 {\cm},
assigned to {\D} totally-symmetric modes (marked by an asterisk
in figs. 1 and 2). The pressure
dependence of their intensity, divided by the intensity of the
normally IR active {\D} $b_u \nu_{51}$ band (at 1285 {\cm}), are reported
in the lower panel of fig. 3. As previously reported,\cite{aoki,noi}
at ambient conditions the stack is already dimerized
to some extent, as shown by the presence of {\it e-mv} induced bands
in the ambient pressure IR spectrum.
By increasing the pressure, the intensity of these bands 
increases gradually, indicating an increase in the 
extent of dimerization.
We remark that by increasing the pressure also 
several Raman bands assigned to totally-symmetric 
modes of {\D} gain appreciable intensity over the bands of {\C}.
This unexpected behavior is clearly evident by 
looking at the pressure evolution of the already mentioned 
{\D} $a_g \nu_{10}$ and $\nu_{5}$ modes in fig.1. A possible
explanation can be formulated on the basis of the ionicity
variation under pressure.
In fact, by increasing the degree of ionicity of {\CD},
either by lowering temperature or by increasing pressure,
a band due to the lowest intra-molecular electronic excitation of {\D}$^-$ 
appears around 2.4 eV in the visible absorption spectra.\cite{aoki}
Since the Raman exciting line used in the present work
is fully in resonance with this band ($\lambda$ = 514.5 nm = 2.41 eV),
\begin{figure}[ht]
\includegraphics* [scale=0.64] {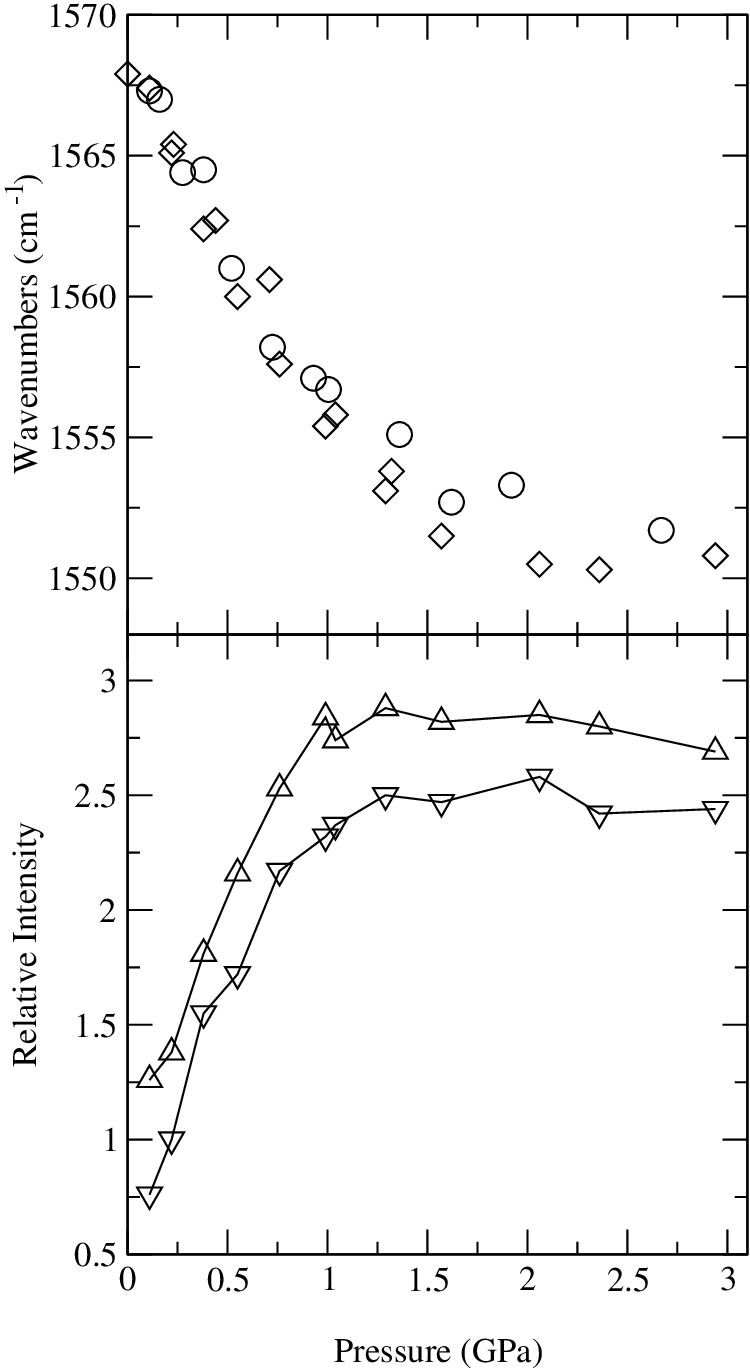}
\protect\vskip0.3truecm
\caption{Upper panel: Pressure dependence of the frequency
of the {\D} $b_u \nu_{46}$ mode. Diamonds and circles
refer to nujol and perflurocarbon spectra, respectively.
Lower panel: Pressure dependence of the relative intensities
of the {\it e-mv} induced bands corresponding to the {\D} $a_g
\nu_5$ (triangles) and $\nu_{10}$ modes (inverted triangles).}
\end{figure}
\noindent
we observe resonance intensity enhancement of the Raman bands due 
to {\D} molecular units.

Before discussing in detail the
pressure induced NI phase transition in {\CD}
one last observation on vibrational spectra is in order.
On the basis of the vibronic model\cite{40years,bibbia} 
for dimerized mixed stack CT crystals, 
{\it e-mv} induced IR bands and their Raman counterparts 
should display frequency coincidence. 
On the contrary, we observe appreciable frequency differences
in the high pressure spectra, the most noticeable one being
observed for the {\D} $a_g \nu_{5}$ band,
which is present at 1585 {\cm} in the Raman 
spectra at 1.53 GPa and at 1598 {\cm} in the IR spectra
at 1.57 GPa (fig. 1). 
The origin of this effect might be again attributed
to the above mentioned resonance Raman conditions. Resonance
effects on Raman spectra under pressure can indeed single out
domains of different ionicities within the crystal, so that
the resulting peak maxima may differ from those of the IR spectra.
Furthermore, one has to keep in mind
that different experimental conditions have been used in measuring
Raman and IR spectra, namely single crystal for Raman, and powdered
samples for IR.

\section{Discussion}

We first discuss the ionicity change of {\CD} crystal under
pressure. Assuming that the frequency of the {\D} $b_u \nu_{46}$
varies linearly with the degree of ionicity,\cite{kobayashi}
from the upper panel of fig. 3 we immediately
derive the pressure dependence of $\rho$ reported
in the left panel of fig. 4. The ionicity increases smoothly
with pressure from $\sim 0.35$ at ambient conditions to $\sim 0.65$
at 3 GPa. In fig. 4, we have limited the pressure to 2.2 GPa, as the frequency
of the $b_u \nu_{46}$ band saturates above such pressure (fig. 3, upper panel).
We remark that to estimate $\rho$ we
have used the frequency of the most intense among the A,A' bands
of fig. 1, so that fig. 4 reports the ionicity of the prevailing phase,
and the possible staging of the transition is not evidenced.
The difference in frequency between A and A' bands is of
the order of 4 {\cm}, corresponding to a $\rho$ difference of the
order of 0.07. The overall ionicity variation is about 0.3
in the pressure range from ambient to 2 GPa. We note that this value 
should be considered a lower limit, as we have not corrected
for the general frequency increase due
\vskip 0.6truecm
\begin{figure}[ht]
\includegraphics* [scale=0.60]{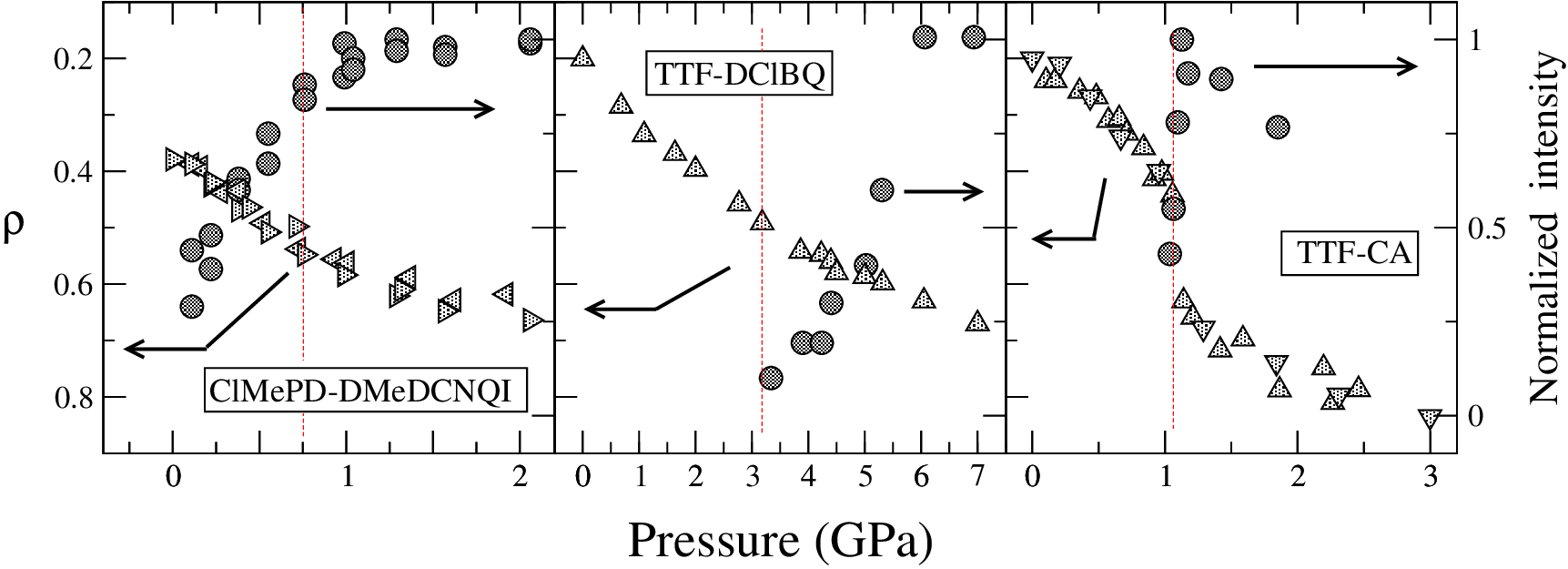}
\protect\vskip0.3truecm
\caption{Pressure dependence of the ionicity ($\rho$)
and of the normalized intensity of the {\it e-mv}
induced bands of {\CD} (left panel), TTF-DClBQ
(center panel, from ref.\protect\onlinecite{ttf25bq})
and TTF-CA (right panel, adapted from refs.\protect\onlinecite{ttfcap,ttfcaj}).
The dashed vertical line is a visual aid to approximately separate
N and I states ($\rho$ between 0.5 and 0.6).}
\vskip 0.2truecm
\end{figure}
\noindent
to compression effects. We have
increased the pressure up to 5.9 GPa, confirming that the
ionicity apparently saturates around 2 GPa. However, above
3 GPa the uncertainty in the estimate
of $\rho$ increases, as the frequency hardening due to
anharmonicity cannot be disregarded anymore.\cite{ttf25bq}

In the left panel of fig. 4 we also report the pressure
dependence of the intensity of
the {\it e-mv} induced IR bands corresponding to the {\D} $a_g~\nu_5$ 
and $\nu_{10}$ modes (fig. 3, lower panel). For convenience,
we have normalized the intensities of both bands to the value at 3 GPa. 
From the figure it is seen that the intensity of these bands,
already present at ambient conditions,  
gradually increases by increasing pressure, reaching a plateau
at about 1.3 GPa. The maximum slope is around 0.5 GPa.
As in the temperature induced phase transition,\cite{noi}
the dimerization occurs well before the crossing of
the NI borderline. On the other hand, the
already mentioned simultaneous presence of domains of different
ionicities (staging in the phase transition)
suggests that the pressure induced phase transition occurs
with a mechanism different from that of the temperature induced
phase transition, which is a truly continuous and
second order transition.
The peculiarity of the NI transition in {\CD} is that the coupling between 
a first order phase transition (order parameter: the degree of ionicity) 
and a second order, displacitive one (order parameter: the degree of 
dimerization) is particularly evident.\cite{noi}

The above interplay between first and second order
phase transition is well appreciated if we compare the pressure
induced NI transition in {\CD}

\vspace {11truecm}
\noindent
with the analogous transitions observed in TTF-DClBQ and in TTF-CA. 
Fig. 4, center and right panel, reports the corresponding
pressure variation of the ionicity, and the normalized intensity of 
the {\it e-mv} induced bands.\cite{ttf25bq,ttfcap,ttfcaj}
In TTF-DClBQ we have a continuous ionicity change, as in {\CD},
but the onset of dimerization occurs when $\rho$ is above
0.5, and the NI borderline has been crossed. In TTF-CA,
on the other hand, the NI transition is first
order, with a ionicity jump of about 0.2 at 1.1 GPa,
and again the stack dimerization sets up once the
NI borderline is crossed.
In both cases, therefore, the transition appears to be driven by 
the degree of ionicity, followed by the dimerization transition. 
In the case of {\CD}, on the opposite, the dimerization starts 
well before the crossover from neutral to ionic regime, causing
a strict interplay
between dimerization and ionicity.
It is worthwhile to remark that in TTF-CA the first order
pressure induced phase
transition is preceded by a precursor regime (from about 0.7 GPa,
not shown in fig. 4),
where dimerized quasi-ionic species are present at the same time
as the quasi-neutral regular stack.\cite{ttfcaj} Such pre-transition
regime is not present in the temperature induced transition
of TTF-CA, at least when well grown single crystals
are used.\cite{ttfcajcp,tokura85} The pre-transition regime has
been interpreted as characterized by the presence of
thermally fluctuating collective low-energy excitations.
However, one cannot exclude that the observed phenomena are
instead due to macroscopic, static domains
of different ionicities, caused by defects in the powdered
samples, as well as by pressure inhomogeneities. The problem is
very similar to that encountered in {\CD}, where we observe
the coexistence of domains of slightly different ionicities.
Such coexistence has not been detected in TTF-DClBQ.\cite{ttf25bq}

The quite different behavior of the pressure induced NI
phase transitions in {\CD}, TTF-DClBQ, and TTF-CA, as
evidenced in fig. 4, can be rationalized on the basis
of existing models of the NI interface. At the basic level,
N or I ground states arise from the competition between $\Delta$,
the energy difference between donor and acceptor sites,
and the energy gain due to the increase of Madelung 
energy following a lattice contraction. The hopping
integral, $t$, mixes the fully N and I quantum states,
and $\rho$ may then change continuously from zero to one.
Electron-electron interactions and {\it e-mv} coupling,
on the other hand, favor first order transitions, with
discontinuous jump of ionicity, whereas the modulation
of $t$ by the intermolecular, lattice phonons ({\it e-lph}
coupling) is able to induce the dimerization of the
stack, yielding a second order, Peierls-like, phase transition.\cite{VB}

Most of the known NI transitions exhibit a discontinuous
jump of ionicity, accompanied by the stack dimerization.
For this reason the interplay between ionicity
and dimerization has been often overlooked. On the other hand, it
has been shown that whereas on the I side the system is
intrinsically unstable towards stack dimerization,
the dimerization transition may also occur on the
\begin{figure}[ht]
\includegraphics* [scale=0.5]{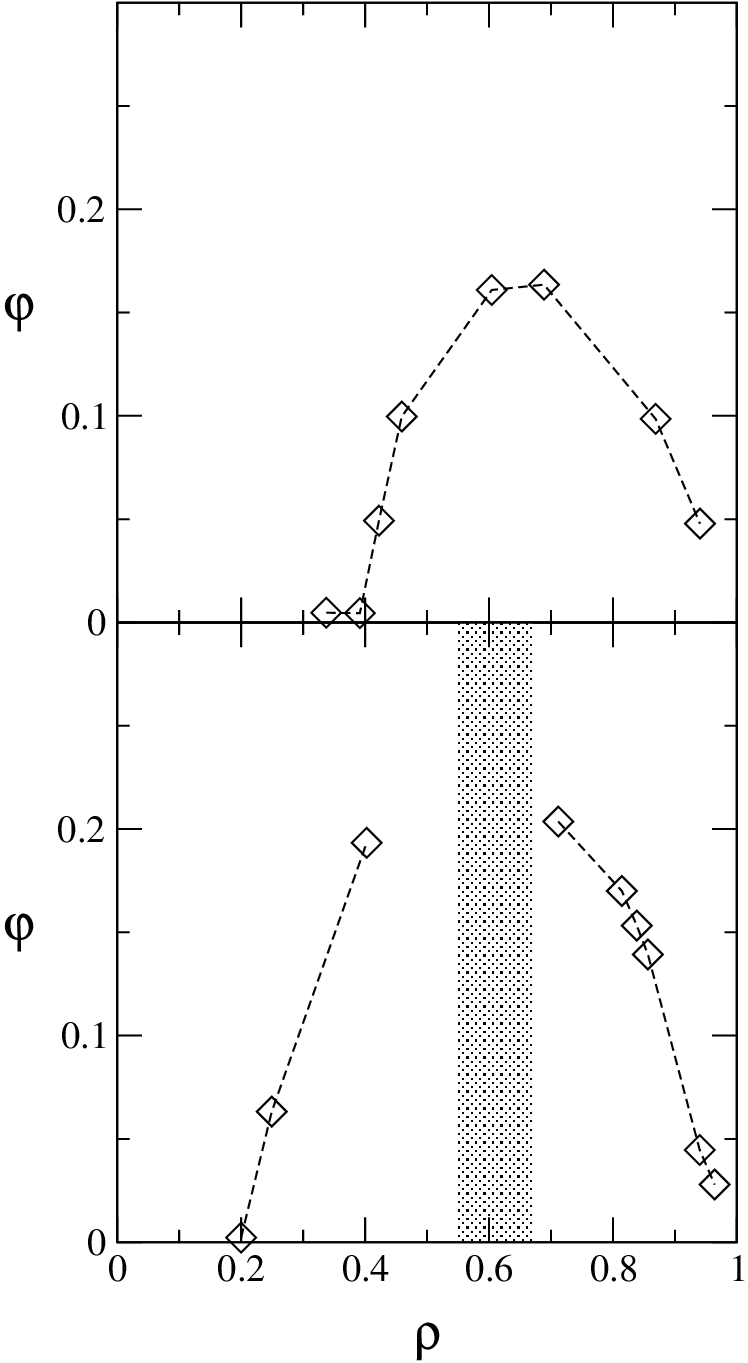}
\protect\vskip0.3truecm
\caption{Asymmetry parameter $\phi \propto(t_{i} - t_{i+1})/(t_{i} + t_{i+1})$
as a function of $\rho$. The calculation is made for the same
set of parameters, except for the electron-electron interaction
strength, which is zero in the lower panel and different
from zero in the upper panel. From ref. \protect\onlinecite{VB}.}
\end{figure}
\noindent
N side, provided the {\it e-lph} coupling is strong enough.\cite{VB} 
The extent of dimerization and the corresponding energy gain
are connected to the ionicity. Fig. 5 shows the calculated
$\rho$ dependence of the asymmetry parameter,
$\phi \propto (t_{i} - t_{i+1})/(t_{i} + t_{i+1})$, directly related
to the extent of dimerization.\cite{VB} The upper panel
refer to a set of parameters that yields a continuous
ionicity change, whereas in the lower panel the
value of the electron-electron interaction gives rise
to a jump in $\rho$ (the shaded area represents a
forbidden ionicity region). TTF-DClBQ and TTF-CA then
can be roughly described by the upper and lower panel of fig. 5,
respectively. In both cases the {\it e-lph} coupling
strength is not enough to induce the dimerization
phase transition before the system is passed on the
I side. Thus the transition is driven by
the ionicity change, and the dimerization follows.
In the case of {\CD}, instead, the {\it e-lph}
coupling yields the stack dimerization on the
N side, when $\rho$ is of the order of 0.35 (fig. 4).
Fig. 5 shows that on the N side $\phi$ increases with $\rho$,
so that the dimerization represents the order parameter
of the second order phase transition, but the $\rho$
increase yields in turn to the increase of $\phi$.
The maximum extent of dimerization, and correspondingly
the maximum gain in energy upon dimerization, occurs
when $\rho$ is between 0.6 and 0.7, and indeed in {\CD}
the pressure dependence of the ionicity
shows tendency to saturation (fig. 4). 

The above sketched scenario of pressure driven
transition in {\CD} is rather simplified,
as it does not consider several factors
which are currently believed important
ingredients of NI phase transitions.
Indeed disorder, low-lying collective
excitations (CT strings),\cite{koshihara1999}
or inter-chain interactions,\cite{collet01,oison01} are not taken into account.
Despite of that, we believe our picture catches the salient
features of NI phase transition in {\CD}, whereas some aspects
still remain to be clarified. From this point of
view, the discovery of a new type of NI transition,
such as that of {\CD}, where the ionicity change is
continuous {\it and} is preceded by the stack dimerization,
represents an additional opportunity for investigating the
fascinating multiple facets of pressure and temperature
induced NI phase transitions.

\acknowledgements
We acknowledge
helpful discussions with A. Painelli.
This work has been supported by the Italian National
Research Council (CNR) within its ``Progetto Finalizzato
Materiali Speciali per tecnologie Avanzate II'', and the
Ministry of University and of Scientific and Technological
Research (MURST).


\end{document}